\documentclass[sigconf]{acmart}
\usepackage[utf8]{inputenc}
\usepackage{graphicx}
\usepackage{amsmath}
\usepackage{caption}
\newtheorem{definition}{Definition}

\AtBeginDocument{%
  \providecommand\BibTeX{{%
    \normalfont B\kern-0.5em{\scshape i\kern-0.25em b}\kern-0.8em\TeX}}}

\copyrightyear{2021}
\acmYear{2021}
\setcopyright{acmcopyright}\acmConference[ICoMS '21]{2021 4th International Conference on Mathematics and Statistics}{June 24--26, 2021}{Paris, France}
\acmBooktitle{2021 4th International Conference on Mathematics and Statistics (ICoMS '21), June 24--26, 2021, Paris, France}
\acmPrice{15.00}
\acmDOI{10.1145/3475827.3475838}
\acmISBN{978-1-4503-8990-7/21/06}

\begin{document}

\title{A Graph Approach to Simulate Twitter Activities with Hawkes Processes}

\author{Ao Qu}
\affiliation{%
  \institution{Vanderbilt University}
  \streetaddress{2201 West End Ave}
  \city{Nashville, TN}
  \country{USA}}
\email{ao.qu@vanderbilt.edu}

\author{Ismael Lemhadri}
\affiliation{%
  \institution{Stanford University}
  \streetaddress{450 Serra Mall}
  \city{Stanford, CA}
  \country{USA}}
\email{lemhadri@stanford.edu}

\begin{abstract}
 The rapid growth of social media has been witnessed during recent years as a result of the prevalence of the internet. This trend brings an increasing interest in simulating social media which can provide valuable insights to both academic researchers and businesses. In this paper, we present a step-by-step approach of using Hawkes process, a self-activating stochastic process, to simulate Twitter activities and demonstrate how this model can be utilized to evaluate the chance of extremely rare web crises. Another goal of this research is to introduce a new strategy that implements Hawkes process on graph structures. Overall, we intend to extend the current Hawkes process to a wider range of scenarios and, in particular, create a more realistic simulation of Twitter activities by incorporating the actual user status and following-follower interactions between users. 
\end{abstract}



\maketitle

\section{Introduction}
Twitter is a micro-blogging platform that allows people to quickly send short
messages, called tweets, on the internet. This platform became very popular, reaching 500 million users in February 2012 \cite{reach500million}. On average, about 6000 tweets are published on Twitter per second, which has triggered an increasing number of studies about how to build a good simulation of twitter traffic \cite{point2015}\cite{log2015}. In particular, Hawkes process, a self-activation process where previous events can have impacts on later occurrences, gives a reasonable way to conduct the simulation. However, previous research lacks visualization of the detailed generating process which makes it hard for practitioners to imagine what is achieved by the model \cite{longitudinal2017}. In this research, we show a step-by-step approach of applying Hawkes proess to the Twitter simulation. Simulating social media offers important insights that can be leveraged to enhance marketing, predicting web-crisis, analyzing information transmission, and adjusting trading strategies \cite{media2016marketing}\cite{media2019stock}. In this paper, we emphasize an important application--using importance sampling to simulate the chance of "Twitpocalypse", a bug that happened a few years ago due to extremely high volume of tweets during a given period of time. However, for other applications mentioned above, it's hard to extract relevant information from Hawkes process model because the model doesn't reflect the specific activities on a micro-scale even under our step-by-step implementation strategy. Some research uses agent-based modeling that considers a specific set of users in the model and makes them carry out certain activities following some fixed rules \cite{media2016marketing}. This approach provides valuable insights but lacks mathematical rigorousness and evaluating numeric results becomes impossible. We realize that one important thing that Hawkes process fails to incorporate is the network structure of social media where the way users are connected can have a huge impact on the performance of the simulation model. Therefore, we design a modified version of Hawkes process which embeds this traditional model within a graph structure that represents the user relationship. This approach successfully combines the advantages of both agent-based modeling which focuses on individual level dynamics and Hawkes process which treats the entire social media as a whole. This modification provides traditional Hawkes process with a broader range of application while maintaining its original self-activating property. The detailed implementation of this model is also discussed in this article. We have not yet conducted a detailed analysis of the statistical properties of the model, which we leave to future work.

\section{Hawkes Process}
\subsection{Model description and implementation}In previous studies, people have been relying on Hawkes process, a self-exciting process brought up by Hawkes, A. G. in 1971, to simulate activities that involve interactions between events \cite{1056305}. For example, an interesting application is to use Hawkes process to model the queues in front of nightclub \cite{daw2018queues}. Usually, we use Poisson process to model the arrivals of customers, or night-club visitors in this case. However, the Poisson model ignores the fact that people may have stronger interests in night-club that seem to be popular(i.e a long queue outside the nightclub). Hawkes process successfully remedied this issue by taking into account the mutual effects between different events. The original formulation of this process is defined as following: \\
\textit{A self-exciting temporal point process N whose conditional intensity function $\lambda = \lambda(t)$, given a specific time t, is defined to be} $$\lambda(t) = \mu(t) + \sum_{i: \tau_i < t} v(t-\tau_i)$$
\textit{where the constant function $\mu(t)$, also denoted as $\lambda_0$ as $\lambda(t) = \mu(t)$ when $t = 0$, is the initial rate of the process $N$ which is usually represented as a constant function, $\tau_i$ are the points in time occurring prior to time $t$, and $v$ is a monotone decreasing non-negative function which governs the clustering density of $N$.} \\

\begin{figure}[hbt!]
\centering
\includegraphics[width=0.45\columnwidth]{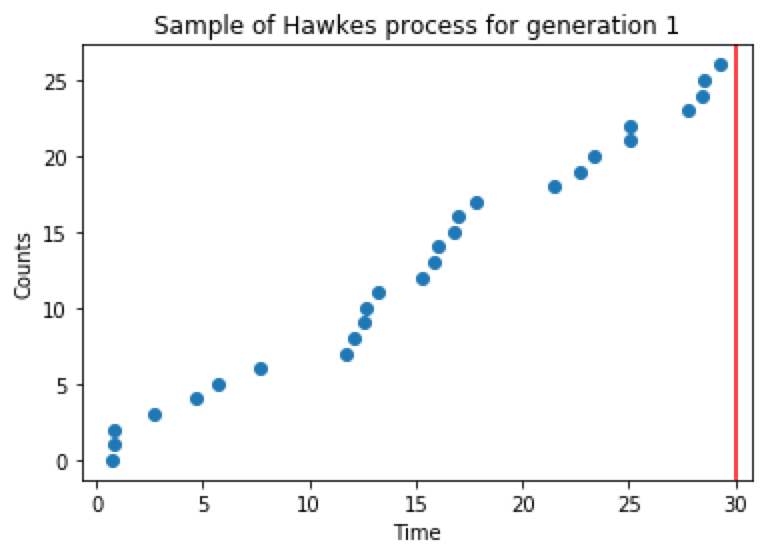}
\includegraphics[width=0.45\columnwidth]{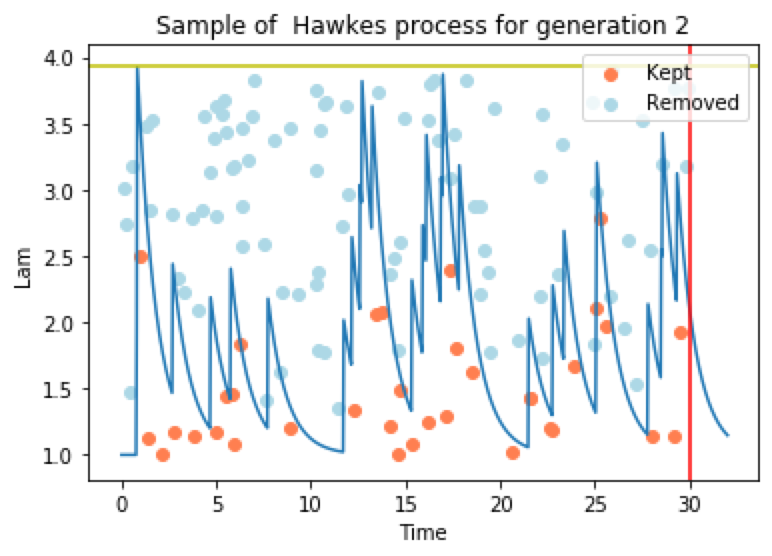}
\includegraphics[width=0.45\columnwidth]{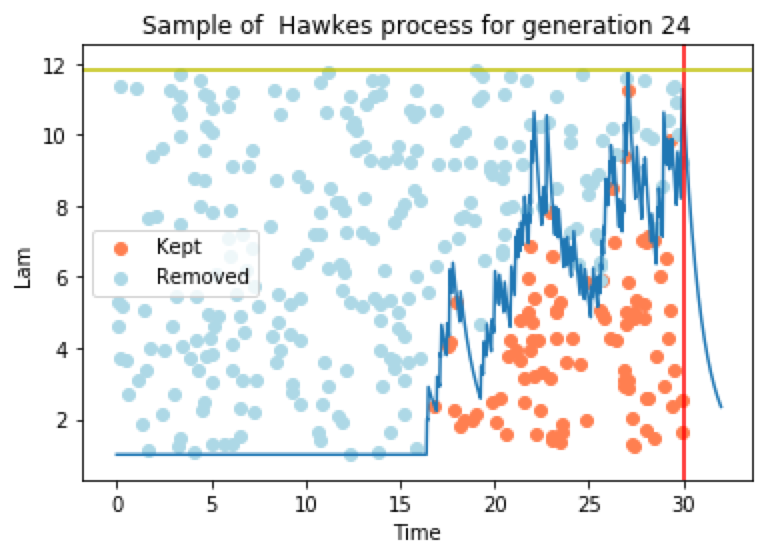}
\includegraphics[width=0.45\columnwidth]{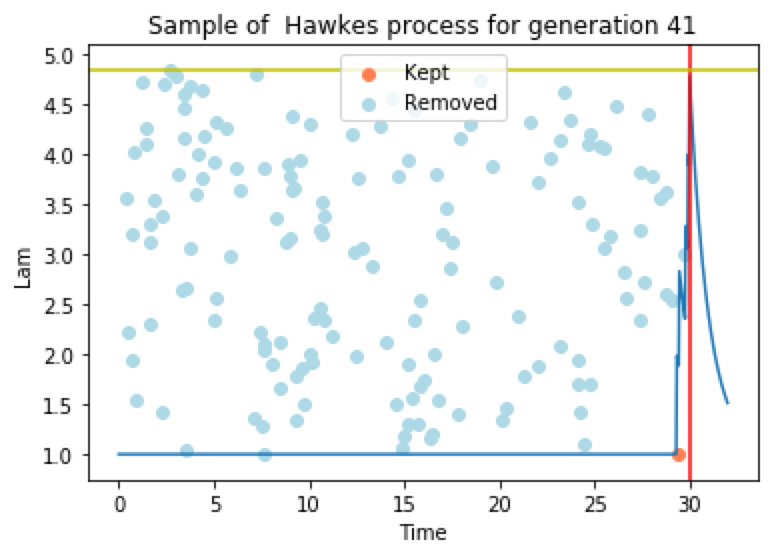}
\caption{Visualization of Hawkes Process}
{\textit{At the beginning, we set the threshold and parameters based on which the first generation is produced. For each generation, we use the maximum $\lambda$ to generate a sample and keep those whose value, as determined by the random number generator, fall under $\lambda(\tau)$. We can see that the number of points first goes up and then goes down as we produce more generations. Eventually, the impact inserted by the early events diminishes and the curve shifts to the right until no more event is generated. }}
\label{hawkesprocess}
\end{figure}

This model has also been implemented to generate simulations of social media activities including tweets \cite{popularity2015}. Intuitively, Hawkes process should be able to produce an accurate simulation of Twitter activities due to its self-exciting property. However, in this article, we will discuss some of the current limitations and bring up a potential solution. We also found that most of the previous studies were more focused on the theoretical part of this model and the actual application was discussed but not explicitly carried out \cite{rizoiu2017tutorial}. In this part, in order to present how this model actually works in application, we write a simulation program using Python with the help of Numpy. To show more details about how the model could be generated step by step, we also visualized the entire process using Matplotlib with relatively small parameters due to the limited computational resources. This example should provide those who are not familiar with Hawkes process or are interested in how the self-exciting property is displayed in practice a comprehensive overview in a way that is easy to understand. \\
\textbf{Step 1: Generate the first generation.} In order to make the calculation more convenient, we implement the Hawkes process generation by generation. For the first generation $k=0$, we simulate times using the homogeneous Poisson process with initial intensity function $\mu: \mathbb{R} \to \mathbb{R}$ defined as $\mu(x) = \lambda_0$ if $x>0$ and $0$ otherwise.
$$P=\{t_{1}^{(0)},...,t_{N_{t}^{(0)}}^{(0)}\}\sim PP(\lambda_{0}),$$ 
 (Poisson process of intensity $\lambda_{0}$)\\
\textbf{Step 2: Define the lambda function.} The definition above has already explained the most part of the model formulation except the function $v$, which is supposed to be a monotone decreasing non-negative function. This function reflects how the impact of previous events diminishes over time but remains positive. In our example, we just follow the convention of using exponential function to represent such trend. We will set parameters $a, b\in \mathbb{R}^{+}$ and define the function $v$: $\mathbb{R}^{+} \to \mathbb{R}^{+}$ as $v(x) = a\cdot e^{-bx}$. \\
\textbf{Step 3: Calculate the maximal intensity for each generation and produce the next generation based on this maximal intensity}. To continue generating the other generations, we shall find the greatest value that $\lambda(t)$ can take given the current generation and use that intensity to generate the next generation. Then, we shall select some events generated during this process to be removed based on how likely these events would happen with the real non-homogeneous intensity. Suppose we have the current generation of $n$ events $\{\tau_i\}_{i=1}^{n}$. Since, the previous generation is a finite set, the maximum exists for this function. Since the function $\lambda(t)$ is decreasing within each interval $[\tau_i, \tau_{i+1}]$ with $i = 1, 2, \dots, n-1$, $max\{\lambda(t): t\in \mathbb{R}^{+}\} \in \{\lambda(\tau_i): i\in 1, 2, \dots, n\}$. Thus, we just need to compare a few values to get the maximum, which significantly reduce the amount of computation.  \\
\textbf{Step 4: Decide which points in the current generation should be kept using random number generator}. It's obvious that following Step 3, we would generate more events than we expected to have because the intensity function doesn't always stay at its maximum. Given a time $\tau$ generated based on the current generation, to decide if we should keep it, we calculate the chance of that event by measuring the ratio of $\lambda(\tau)$ to $max\{\lambda(t): t\in \mathbb{R}^{+}\}$ which has already been determined. When we finalize the $k^{th}$ generation, we will denote the set of events as $P_k$ and move on to generate the $k+1^{th}$ generation using the same procedure.\\
\textbf{Step 5: Terminate the algorithm and build the superimposed process} For each generation, we will set a time threshold $T$ so that events generated after $T$ will be discarded as time dimension can go to infinity and we won't keep track of it. If there is no new event generated based on the current generation, we will decide to terminate the algorithm as the previous impacts have shrunk to minimal. The randomness of this process makes the number of generations hard to predict. However, some previous studies have shown that we can compute the expected value and standard deviation with the help of differential equations. We shall also experimentally validate those results in our current study. Then, the entire Hawkes process can be summarized by summing up all the previous generations
$$\{t_{1},..,t_{N_{T}}\}=\bigcup_{k=0}^{K-1}\Pi_{(k)}$$
where $N_{t}=\sum_{k=0}^{K-1}N_{t}^{(k)}.$ \\
\textbf{Step 6: Visualization} The simulation result can be visualized using Python. Here, as shown in Figure \ref{hawkesprocess}, we pick two generations that clearly demonstrate how the events are produced, generation by generation. 

\subsection{Complexity Analysis}
\begin{figure}[hbt!]
\centering
\includegraphics[width=0.8\columnwidth]{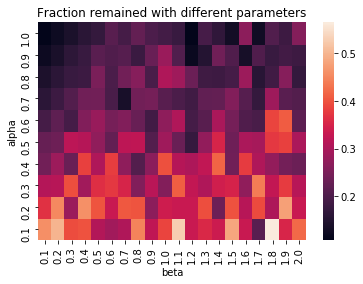}
\caption{Heatmap of Model's Complexity}
{\textit{We run experiments with different sets of parameters and output the results in this heat map visualization. We can see that as $\alpha$ gets larger and $\beta$ gets smaller, the efficiency of our algorithm drops as less points generated are used as the final output. Therefore, our method might fit some specific scenarios but we should seek better approaches when dealing with larger tasks.}}
\label{heatmap}
\end{figure}

The efficiency of our implementation depends on the ratio between the number of actual events that are accepted and the number of total events generated. As shown in Figure \ref{heatmap}, this algorithm works well with small sample size and small parameters but gets less efficient with larger parameters and time span. Previously, Ogata(\cite{1056305}) proposed a fast algorithm implementing Hawkes process. Put briefly, their algorithm generates the entire Hawkes process simultaneously instead of breaking it into several generations. Here, we chose to use this step-by-step method for simplicity and in order to clarify the mechanism and make better visualization.

\section{Application: Rare Event Simulation}
One important application of this model is that it can be utilized to estimate the probability of an extremely rare event--"twitpocalypse" \cite{twitpocalypse}. Twitter labeled each tweet with an unique ID and the largest amount of IDs that can be maintained is $2,147,483,647$, the largest number that can be stored as "signed integer". A decade ago, more than 20 billion tweets had been sent under 3 years, which went far beyond the expectation of Twitter developers and caused a crash. Moreover, a lot of third-party apps that were connected with Twitter couldn't handle this large number and also crashed during this incident. As Twitter gains increasing popularity, some people have been suspicious about with what chance there will another "twitpocalypse" and when it will happen. Due to Hawkes Process's superior ability to simulate the nature of social media streams, it can serve as a good tool to approximate the chance of reaching an enormous traffic in social media with the aid of importance sampling. In this section, we will introduce the most common importance sampling method and discuss how this method can be implemented under this Hawkes Process context.
\subsection{Importance Sampling}
In statistics, importance sampling is a popular variance reduction strategy used to estimate a rare event that is hard to be obtained using traditional ways such as direct calculation and Monte Carlo estimation \cite{Glynn96importancesampling}. Most of the distributions are too complicated to derive a formula for computing probability so people usually just use Monte Carlo estimation which runs a large number of experiments and approximates the actual probability of a event with the empirical probability. However, if the actual probability is extremely low, as in the case of Twitpocalypse, we will need a huge amount of experiments just to get one single occurrence of the event and clearly this estimated probability is very unstable. Therefore, when the chance of a certain event is really low, it is a common practice to use importance sampling which reduced the variance and results in a much higher robustness \cite{bucklew_2004}. The formal description of one way to do importance sampling is following:
Suppose we want to estimate $\rho = \mathbb{E}[\eta(X)]$ where $X$ is a random variable describing some observation, $\eta$ is an indicator function of some events, and $\rho$ is just the probability of this set of events. Then, one way to estimate $\rho$ is to generate a sequence of i.i.d. random numbers $X^{(1)}, X^{(2)}, \dots, X^{(n)}$ and then compute the empirical probability
$$\hat{\rho} = \frac{1}{n}\sum_{i=1}^{n} \eta(X^{(i)}).$$
However, as we just discussed, this could lead to an inaccurate and unstable result if the probability we are estimating is too low. In order to solve this, we can introduce a new variable variable $Y$ and generate a sequence of i.i.d. random numbers $Y^{(1)}, Y^{(2)}, \dots, Y^{(k)}$. Suppose $X$ is equipped with a probability density function $p(\cdot)$ and $Y$ is equipped with a probability density function $q(\cdot)$. Then, as long as $support(p\cdot \eta) \subset support(q\cdot \eta)$, we can show
\begin{align*}
    \hat{\rho} &= \frac{1}{n}\sum_{i=1}^{n} \eta(X^{(i)}) \\
    &= \frac{1}{k}\sum_{i=1}^{n} \int \eta(x^{(i)}) p(x^{(i)}) dx \\
    &= \frac{1}{k} \sum_{i=1}^{n} \int \eta(x^{(i)}) \frac{p(x^{(i)})}{q(x^{(i)})} q(x^{(i)}) dx \\
    &= \mathbb{E}_{q} [\eta(Y) \frac{p(Y)}{q(Y)}] \\
    &= \frac{1}{n}\sum_{i=1}^{n} \eta(Y^{(i)})\frac{p(Y^{(i)})}{q(Y^{(i)})} \\
\end{align*}
In practice, we can define a random variable $Y$ where the rare even is much more likely to happen and run a Monte Carlo simulation with this $Y$. This result has been proved to be unbiased. In the next section, we will discuss the specific implementation of this technique in Hawkes Process and how the probability of a rare event, "Twitpocalypse", can actually be estimated.

\subsection{How to implement on Hawkes Process}
\begin{figure}
\centering
\includegraphics[width=0.7\columnwidth]{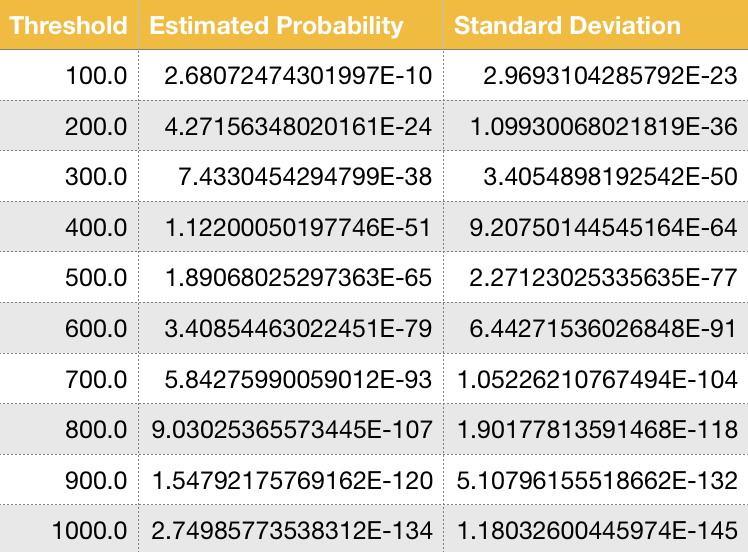}
\caption{Estimating the chance of Twitpocalypse with Importance Sampling}
{\textit{We use Monte Carlo simulation technique to estimate the probability of exceeding each of these $10$ thresholds within $10$ time units, each with $100$ experiments. Then, we calculate the estimated probability and standard deviation for each threshold. We can see that as the threshold gets larger, the chance of exceeding that threshold within a given time period gets extremely low. Also, we can tell that the model is very stable because the standard deviation is about 12 to 13 orders of magnitude smaller than the expected value.}}
\label{numericalresults}
\end{figure}

For Hawkes Process, previous research has derived several theorem regarding its limit behavior, expectation, and variance of the intensity and number of events generated. In this paper, we will demonstrate how we can put together the previous results and apply them into an importance sampling implementation. Given a certain set of parameters $\alpha, \beta, \lambda_0 = \mu(t)$, the limit of intensity $\lambda_t$ as $t\to \infty$ is formulated as $$\lambda_{\infty} = \frac{\beta \lambda_0}{\beta - \alpha}.$$
Also, the formula for approximating the expected value has been obtained by solving a differential equation system (\cite{daw2018queues}),
\begin{equation}
  \mathbb{E}[N_t] = 
    \begin{cases}
    \lambda_{\infty}t + \frac{\lambda_0 - \lambda_{\infty}}{\beta - \alpha}(1 - e^{-(\beta-\alpha)t})   & \text{if $\alpha<\beta$}\\ \\
    \bigg(\frac{\beta \lambda_0}{(\alpha-\beta)^2} + \frac{\lambda_0}{\alpha-\beta}\bigg) (e^{(\alpha-\beta)t} - 1) - \frac{\beta \lambda_0}{\alpha-\beta}t & \text{if $\alpha>\beta$}\\ \\
    \frac{\beta \lambda_0}{2} t^2 + \lambda_0 t  & \text{if $\alpha=\beta$}
    \end{cases}       
\end{equation}
From these formula, it's clear that as we increase the initial intensity $\lambda_0$, the expected number of events will increase. Therefore, our strategy is to build another Hawkes Process with a larger initial intensity $\hat{\lambda_0}$ such that the corresponding expected value is equal to the extreme value we are trying to simulate. In this case, the extreme value is just the maximum number of tweets that can be stored as a "signed integer" and handled by the system, which is approximately equal to $2^{32}$. Hence, the occurrence of this rare event can be represented as $N_t> 2^{32}$ with $t$ being the time period we are interested in. Therefore, we can first compute $\mathbb{E}[N_t | \lambda_0]$ and set $$\hat{\lambda_0} = \frac{2^{32}}{\mathbb{E}[N_t | \lambda_0]}.$$
The next step is to compute the ratio between the likelihoods of these two processes to fully construct the importance sampling. 
Another research has shown that the likelyhood function for Hawkes Process takes the following form: 
\textit{
Suppose our Hawkes Process takes $\lambda_0, \alpha, \beta$ as its parameters and $t$ as its time threshold. Then, by using results on likelihood of point process, we can obtain the likelihood of this Hawkes Process as 
$$L_{\lambda_0} = exp\{ \int_{0}^{t} (1-\lambda(s)) ds + \int_{0}^{t} ln(\lambda(s)) dN_s  \}.$$
In particular, if we have the collection of tweets $\{t_1, t_2, \dots, t_n\}$, the above equation can be rewritten as
$$L_{\lambda_0}(t_1, t_2, \dots, t_n) = exp\{t - t\lambda - \sum_{i=1}^{n} \alpha (1 - e^{-\beta(t - t_i)}) + \sum_{i=1}^{n} ln(\lambda(t_i)) \}.$$
}
Using the importance sampling technique mentioned in the last section, if we take a very large $m$ and run the experiments with the Hawkes Process with initial intensity $\hat{\lambda_0}$ as discussed above. We can use the results, denoted as $Y_1, Y_2, \dots, Y_m$ where each $Y_k$ is a set of all the tweets $t_{k,1}, t_{k, 2}, \dots, t_{k, n}$ generated in that experiment, to estimate the actual probability of original Hawkes Process generating more than $2^{31}$ tweets within $t$ time period
$$\hat{\rho} = \frac{1}{m}\sum_{i=1}^{m} \eta(Y^{(i)}) \frac{L_{\lambda_0} (Y^{(i)})}{L_{\hat{\lambda_0}} (Y^{(i)})}.$$
Note: in this case, $\eta(Y^(i))$ is the indicator variable about if $i^{th}$ experiment produces more than $2^{32}$ tweets and this should appear very frequent as we have carefully chosen a large $\hat{\lambda_0}$.
Estimating the real probability of twitpocalypse would require significant amount of computational resources due to the large sample size. In our research, we evaluate the probability of generating more than $1000$ tweets within $10$ units of time which, according to the formula presented above, is supposed to have an expected value of $34.548$. The numeric results are summarized in Table \ref{numericalresults}.

\section{Limitations of the traditional approach}
However, the traditional approach of representing Twitter activities using Hawkes process fails to consider the social network structure and individual differences, which would limit the model's performance in a more micro-scale simulation and prediction. The traditional model may have an important role in predicting extremely rare events like "Twitpocalypse", but the scope of applications is restricted. For example, we may not find the traditional model very helpful if we are interested in questions like "under what kind of social network structure will a given user's tweets be most likely forwarded?" In practice, a lot of tasks that require a simulation of Twitter activities are focused on micro-scale Twitter activity \cite{kobayashi2016tideh} and aiming to explore more detailed properties about Twitter usage \cite{10.1145/1871985.1871993}. By far, although modeling of social network has received an increasing interest due to the data availability and growth in media usage, very few techniques have been developed for modeling social media activities with a focus on the impact brought by a specific network structure. With the same set of users, if we represent their relationships with a network structure, different networks might generate totally different activity patterns. A good property of Hawkes process is that it successfully captures how a series of events interact and affect each other, which can't be represented by the traditional Poisson process due to its memoryless property. Therefore, we intend to incorporate this important property into our model while embedding it with a graph structure so that more detailed information is included. In the next section, we will give a thorough description of the model design and a brief example of its implementation.

\section{Graph Model}
\subsection{Model Definition}
The traditional way of simulating Twitter activities using Hawkes Process considers all the twitter activities as a whole but fails to capture individual difference. This method could be useful in measuring and estimating the total Twitter volume but hard to be applied into some more micro-level scenarios like predicting the number of retweets that a message can generate. At this point, we also realize that the structure of social network plays a significant role in users' twitter activity. For example, it might be easier for a user with more followers to get more retweets.
In this part, we will introduce a more realistic simulation that embeds Hawkes Process into a graph structure and simulate each user's twitter activity(represented by node) based on the arrival of tweets sent by those the user is following. In our model, each node will represent a twitter user and the edges (directed) can represent if one user is following or followed by another user. 

\begin{definition}
Given a group of users, if any of them is not followed or following any user not from this group, this group is called a closed Twitter network.
\end{definition}

\begin{definition}
If one user is followed by or following another user, we say these two users are connected. If a is connected to b and b is connected to c, then a is connected to c. In other words, transitivity holds. Given a closed Twitter network, if every pair of users in this network are connected, then this network is irreducible.
\end{definition}

\begin{definition}

Each closed and irreducible Twitter network with $n$ users has a directed graph representation $G = (V, E)$ with $V = \{V_1, V_2, \dots, V_n\}$ (each vertex represents a user). For each pair of $(V_i, V_j)$ such that $V_i$ is following $V_j$, there is an edge $(V_j, V_i) \in E$.
\end{definition}

\begin{definition}
For each $i$ in $\{1, \dots, n\}$, there is an intensity $I_i$. If $V_i$ doesn't follow anyone else on Twitter, then the number of tweets published by $V_i$ follows a Poisson process with intensity $I_i$. $V_i$'s twitter activity can be affected by the tweets published by all users that $V_i$ is following. Let $U_i = \{V_j: (i, j) \in E\}$, namely the collection of users that $V_i$ is following. Let $X_{i,\tau}$ denote a tweet message published by the user $V_i$ at time $\tau$. Given a period of time $T$, let $M_i$ be the collection of tweet messages such that $M_i = \{X_{k, \tau}: k \in U_i, \tau \in T\}$. Then the retweet activity of $V_i$ during time $T$ can be modeled by a non-homogeneous Poisson process with intensity $\lambda(t) = \sum_{\{\tau: \tau \leq t, \exists k \in U_i X_{k, \tau} \in M_i\}} g(t - \tau)$.
\end{definition}

Instead of making the whole tweet generating process as a Hawkes process, we assign a Hawkes function on each individual that also responds to others' posts. This model inherits the advantage of Hawkes model that takes the impacts of previous events into account. However, the network structure of this model makes it more suitable for social media simulation. Also, with this graph structure, it becomes possible to do some in-depth analysis, such as quantifying the impact generated by a KOL, with the help of some graph theory approaches. 

\subsection{Implementation Specifics}
\begin{figure} 
    \centering
    \vspace{0.5cm}
    \includegraphics[width=0.7\columnwidth]{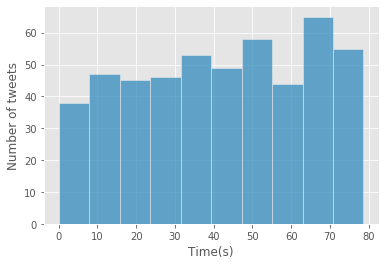}
    \caption{Histogram of tweets generated over time}
    \label{Histogram}
\end{figure}

\begin{figure}
    \centering
    \vspace{0.5cm}
    \includegraphics[width=0.7\columnwidth]{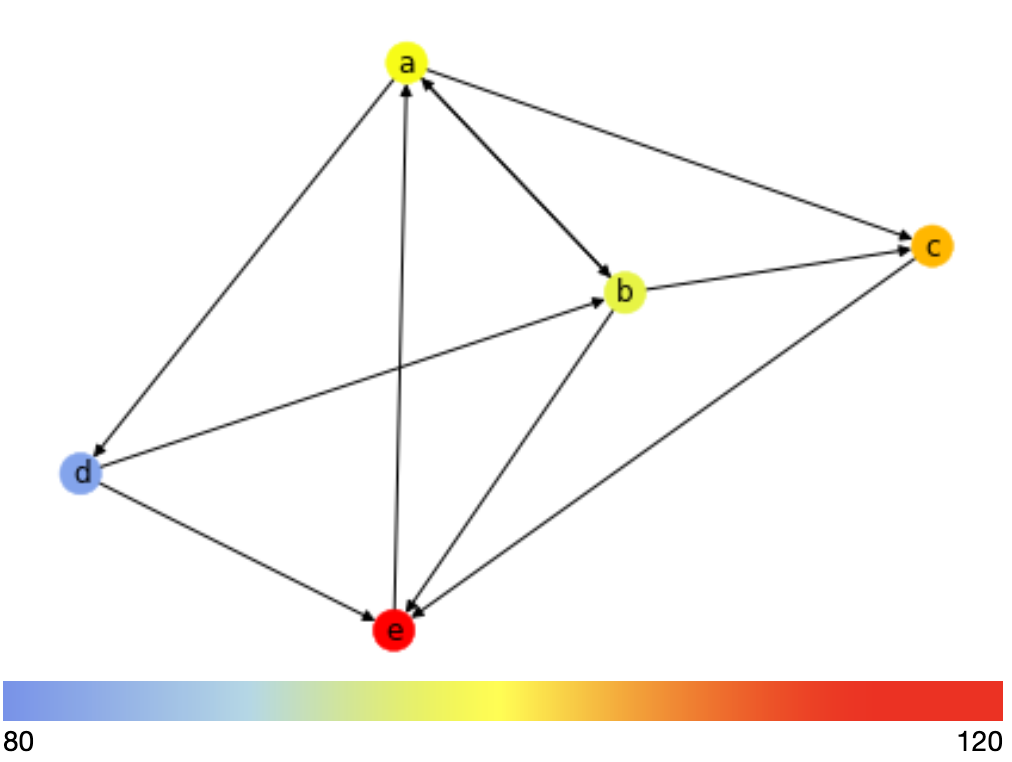}
    \caption{Visualization of network}
    \label{Network}
\end{figure}

The implementation of this model is similar to the original Hawkes process. The major challenge in this model is that the function of each vertex (user) changes according to other vertices. Our approach is that due to the memoryless property of exponential function, instead of generation by generation, we can run the simulation step by step. The function for each vertex is fixed between two consecutive tweets in the whole network. Therefore, we can always use the functions at current step to generate to generate the next tweet for each individual and pick the earliest tweet to be the one that actually takes place. Then, this tweet will help us update the intensity function of all the users who are following this publisher and run the next round of simulation. We experimented on a small network with five vertices. With deliberate choice of parameters, the tweets are generated at a very stable rate (Figure \ref{Histogram}), which is similar to the real world situation. We can also see from Figure \ref{Network} that the user who follows more accounts is likely to produce more tweets.


The flexibility of this model opens up various options for applications. For example, we can assign customized parameters ($\alpha, \beta$) to each vertex to represent different user segments. Also, we can study the gain of adding a particular edge and the loss of deleting a particular edge to measure how a minor change could insert impact on the whole network.

\section{Challenges and Future Work}
The current implementation of the graph model is not as efficient as the traditional Hawkes model. This is because we have to compare the next tweet of each user in order to prepare and update the model for the next step. The complexity of this implementation algorithm could be very high as we have to do $O(|V(G)|)$ times of computations and each computation goes through all the relevant past tweets which stack up over time. A possible simplification strategy is to clean up the previous tweets whose impacts become minimal as they have been published for quite a while so that we can avoid some unnecessary computations. Another challenge is that it is really hard to employ analysis on the limit behavior and other statistical properties such as expected value and variance in this model due to the complicated and frequent interactions between vertices. Our future work will be primarily focused on developing these properties of the model to improve rigorousness and robustness.

\section{Conclusion}
In this paper, we give an overview of the Hawkes process theory and introduce a reproducible strategy to implement it using Python. This step-by-step approach can help keep track of how the events are generated and give researchers opportunities to observe the trend and adjust the model in accordance with the real pattern that is being studied. Also, we compare the complexity of our approach with an existing fast implementation of Hawkes process. On the practical side, we demonstrate an interesting application of Hawkes process that estimates the probability of an extremely rare event known as "Twitpocalypse". However, regarding social media modeling, the scope provided by this traditional Hawkes model is limited because the dynamics of social media activities also depend on the network structure that is formed by the users. In order to address this issue, we bring up a new model which embeds the Hawkes process in a graph structure that incorporates both the self-activating property and individual level features. Due to the complexity of this new model, in this paper, we just briefly illustrate the implementation method and run an experiment on a small sample graph. Although there is still a lot of further work that needs to be done to make this model more rigorous, the potential of this model can be seen in its ability to consider very specific situations and simulate the individual-level interactions.

\bibliographystyle{ACM-Reference-Format}
\bibliography{refs}


\begin{thebibliography}{15}


\ifx \showCODEN    \undefined \def \showCODEN     #1{\unskip}     \fi
\ifx \showDOI      \undefined \def \showDOI       #1{#1}\fi
\ifx \showISBNx    \undefined \def \showISBNx     #1{\unskip}     \fi
\ifx \showISBNxiii \undefined \def \showISBNxiii  #1{\unskip}     \fi
\ifx \showISSN     \undefined \def \showISSN      #1{\unskip}     \fi
\ifx \showLCCN     \undefined \def \showLCCN      #1{\unskip}     \fi
\ifx \shownote     \undefined \def \shownote      #1{#1}          \fi
\ifx \showarticletitle \undefined \def \showarticletitle #1{#1}   \fi
\ifx \showURL      \undefined \def \showURL       {\relax}        \fi
\providecommand\bibfield[2]{#2}
\providecommand\bibinfo[2]{#2}
\providecommand\natexlab[1]{#1}
\providecommand\showeprint[2][]{arXiv:#2}

\bibitem[\protect\citeauthoryear{Bucklew}{Bucklew}{2004}]%
        {bucklew_2004}
\bibfield{author}{\bibinfo{person}{James~A. Bucklew}.}
  \bibinfo{year}{2004}\natexlab{}.
\newblock \bibinfo{booktitle}{\emph{Introduction to rare event simulation}
  (\bibinfo{edition}{1} ed.)}.
\newblock \bibinfo{publisher}{Springer}.
\newblock
\urldef\tempurl%
\url{https://doi.org/10.1007/978-1-4757-4078-3}
\showDOI{\tempurl}


\bibitem[\protect\citeauthoryear{{David Minton}}{{David Minton}}{2009}]%
        {twitpocalypse}
\bibfield{author}{\bibinfo{person}{{David Minton}}.}
  \bibinfo{year}{2009}\natexlab{}.
\newblock \bibinfo{booktitle}{\emph{A user's guide to understanding the
  Twitpocalypse}}.
\newblock
\newblock
\shownote{\url{https://designhammer.com/blog/users-guide-understanding-twitpocalypse}.}


\bibitem[\protect\citeauthoryear{Daw and Pender}{Daw and Pender}{2018}]%
        {daw2018queues}
\bibfield{author}{\bibinfo{person}{Andrew Daw} {and} \bibinfo{person}{Jamol
  Pender}.} \bibinfo{year}{2018}\natexlab{}.
\newblock \bibinfo{title}{Queues Driven by Hawkes Processes}.
\newblock
\newblock
\showeprint[arxiv]{1707.05143}~[math.PR]


\bibitem[\protect\citeauthoryear{Emilio~Serranoa}{Emilio~Serranoa}{2016}]%
        {media2016marketing}
\bibfield{author}{\bibinfo{person}{Carlos A.~Iglesias Emilio~Serranoa}.}
  \bibinfo{year}{2016}\natexlab{}.
\newblock \showarticletitle{Validating viral marketing strategies in Twitter
  via agent-based social simulation}.
\newblock \bibinfo{journal}{\emph{Expert Systems with Applications}}
  (\bibinfo{year}{2016}).
\newblock


\bibitem[\protect\citeauthoryear{et~al.}{et~al.}{2015a}]%
        {point2015}
\bibfield{author}{\bibinfo{person}{J.~R.~Zipkin et al.}}
  \bibinfo{year}{2015}\natexlab{a}.
\newblock \showarticletitle{Point-process models of social network
  interactions: Parameter estimation and missing data recovery}.
\newblock \bibinfo{journal}{\emph{European Journal of Applied Mathematics}}
  (\bibinfo{year}{2015}).
\newblock


\bibitem[\protect\citeauthoryear{et~al.}{et~al.}{2017}]%
        {longitudinal2017}
\bibfield{author}{\bibinfo{person}{P~K~Srijith et al.}}
  \bibinfo{year}{2017}\natexlab{}.
\newblock \showarticletitle{Longitudinal Modeling of Social Media with Hawkes
  Process based on Users and Networks}.
\newblock \bibinfo{journal}{\emph{ASONAM '17: Proceedings of the 2017 IEEE/ACM
  International Conference on Advances in Social Networks Analysis and Mining}}
  (\bibinfo{year}{2017}).
\newblock


\bibitem[\protect\citeauthoryear{et~al.}{et~al.}{2015b}]%
        {log2015}
\bibfield{author}{\bibinfo{person}{Qingyuan~Zhao et al.}}
  \bibinfo{year}{2015}\natexlab{b}.
\newblock \showarticletitle{Modeling Tweet Arrival Times using Log-Gaussian Cox
  Processes}.
\newblock \bibinfo{journal}{\emph{Proceedings of the 21th ACM SIGKDD
  International Conference on Knowledge Discovery and Data Mining}}
  (\bibinfo{year}{2015}).
\newblock


\bibitem[\protect\citeauthoryear{Glynn}{Glynn}{1996}]%
        {Glynn96importancesampling}
\bibfield{author}{\bibinfo{person}{Peter~W. Glynn}.}
  \bibinfo{year}{1996}\natexlab{}.
\newblock \bibinfo{booktitle}{\emph{Importance Sampling For Monte Carlo
  Estimation Of Quantiles}}.
\newblock \bibinfo{type}{{T}echnical {R}eport}.
  \bibinfo{institution}{Publishing House of Saint Petersburg University}.
\newblock


\bibitem[\protect\citeauthoryear{Kobayashi and Lambiotte}{Kobayashi and
  Lambiotte}{2016}]%
        {kobayashi2016tideh}
\bibfield{author}{\bibinfo{person}{Ryota Kobayashi} {and}
  \bibinfo{person}{Renaud Lambiotte}.} \bibinfo{year}{2016}\natexlab{}.
\newblock \bibinfo{title}{TiDeH: Time-Dependent Hawkes Process for Predicting
  Retweet Dynamics}.
\newblock
\newblock
\showeprint[arxiv]{1603.09449}~[cs.SI]


\bibitem[\protect\citeauthoryear{Leung, Wong, and Wong}{Leung
  et~al\mbox{.}}{2019}]%
        {media2019stock}
\bibfield{author}{\bibinfo{person}{Woon~Sau Leung}, \bibinfo{person}{Gabriel
  Wong}, {and} \bibinfo{person}{Woon~K. Wong}.}
  \bibinfo{year}{2019}\natexlab{}.
\newblock \showarticletitle{Social-Media Sentiment, Portfolio Complexity, and
  Stock Returns}.
\newblock \bibinfo{journal}{\emph{Available at SSRN:
  https://ssrn.com/abstract=3492722 or http://dx.doi.org/10.2139/ssrn.3492722}}
  (\bibinfo{year}{2019}).
\newblock


\bibitem[\protect\citeauthoryear{{Ogata}}{{Ogata}}{1981}]%
        {1056305}
\bibfield{author}{\bibinfo{person}{Y. {Ogata}}.}
  \bibinfo{year}{1981}\natexlab{}.
\newblock \showarticletitle{On Lewis' simulation method for point processes}.
\newblock \bibinfo{journal}{\emph{IEEE Transactions on Information Theory}}
  \bibinfo{volume}{27}, \bibinfo{number}{1} (\bibinfo{year}{1981}),
  \bibinfo{pages}{23--31}.
\newblock


\bibitem[\protect\citeauthoryear{Rao, Yarowsky, Shreevats, and Gupta}{Rao
  et~al\mbox{.}}{2010}]%
        {10.1145/1871985.1871993}
\bibfield{author}{\bibinfo{person}{Delip Rao}, \bibinfo{person}{David
  Yarowsky}, \bibinfo{person}{Abhishek Shreevats}, {and}
  \bibinfo{person}{Manaswi Gupta}.} \bibinfo{year}{2010}\natexlab{}.
\newblock \showarticletitle{Classifying Latent User Attributes in Twitter}. In
  \bibinfo{booktitle}{\emph{Proceedings of the 2nd International Workshop on
  Search and Mining User-Generated Contents}} (Toronto, ON, Canada)
  \emph{(\bibinfo{series}{SMUC '10})}. \bibinfo{publisher}{Association for
  Computing Machinery}, \bibinfo{address}{New York, NY, USA},
  \bibinfo{pages}{37–44}.
\newblock
\showISBNx{9781450303866}
\urldef\tempurl%
\url{https://doi.org/10.1145/1871985.1871993}
\showDOI{\tempurl}


\bibitem[\protect\citeauthoryear{Rizoiu, Lee, Mishra, and Xie}{Rizoiu
  et~al\mbox{.}}{2017}]%
        {rizoiu2017tutorial}
\bibfield{author}{\bibinfo{person}{Marian-Andrei Rizoiu},
  \bibinfo{person}{Young Lee}, \bibinfo{person}{Swapnil Mishra}, {and}
  \bibinfo{person}{Lexing Xie}.} \bibinfo{year}{2017}\natexlab{}.
\newblock \showarticletitle{A tutorial on hawkes processes for events in social
  media}.
\newblock \bibinfo{journal}{\emph{arXiv preprint arXiv:1708.06401}}
  (\bibinfo{year}{2017}).
\newblock


\bibitem[\protect\citeauthoryear{{VentureBeat}}{{VentureBeat}}{2012}]%
        {reach500million}
\bibfield{author}{\bibinfo{person}{{VentureBeat}}.}
  \bibinfo{year}{2012}\natexlab{}.
\newblock \bibinfo{booktitle}{\emph{Twitter reaches 500M users, 140M in the
  U.S.}}
\newblock
\newblock
\shownote{\url{https://venturebeat.com/2012/07/30/twitter-reaches-500-million-users-140-million-in-the-u-s/}
  (accessed May 24, 2020).}


\bibitem[\protect\citeauthoryear{Zhao, Erdogdu, He, Rajaraman, and
  Leskovec}{Zhao et~al\mbox{.}}{2015}]%
        {popularity2015}
\bibfield{author}{\bibinfo{person}{Qingyuan Zhao}, \bibinfo{person}{Murat~A.
  Erdogdu}, \bibinfo{person}{Hera~Y. He}, \bibinfo{person}{Anand Rajaraman},
  {and} \bibinfo{person}{Jure Leskovec}.} \bibinfo{year}{2015}\natexlab{}.
\newblock \showarticletitle{SEISMIC: A Self-Exciting Point Process Model for
  Predicting Tweet Popularity}.
\newblock \bibinfo{journal}{\emph{Proceedings of the 21th ACM SIGKDD
  International Conference on Knowledge Discovery and Data Mining}}
  (\bibinfo{year}{2015}).
\newblock


\end{thebibliography}

\end{document}